# *Light beam carrying natural non-integer orbital angular momentum in free space*

Xiaoyu Weng[1]*, Yu Miao[2], Guanxue Wang[2], Qiufang Zhan[2], Xiangmei Dong[2], Junle Qu[1], Xiumin Gao[2]* and Songlin Zhuang[2]

*1 Key Laboratory of Optoelectronic Devices and Systems of Ministry of Education and Guangdong Province, College of Physics and Optoelectronic Engineering, Shenzhen University, Shenzhen, 518060, China.*

*2 Engineering Research Center of Optical Instrument and System, Ministry of Education, Shanghai Key Lab of Modern Optical System, School of Optical-Electrical and Computer Engineering, University of Shanghai for Science and Technology, 516 Jungong Road, Shanghai 200093, China.*

\* Correspondence and requests for materials should be addressed to X.W. (email: xiaoyu@szu.edu.cn) or to X.G. (email: gxm@usst.edu.cn).

# Abstract:

Light beam with optical vortices can propagate in free space only with integer orbital angular momentum. Here, we invert this scientific consensus theoretically and experimentally by proposing light beams carrying natural non-integer orbital angular momentum. These peculiar light beams are actually special solutions of wave function, which possess optical vortices with the topological charge $l+0.5$, where $l$ is an integer. Owing to the interaction of phase and polarization singularity, these vortex beams with fractional topological charge can maintain their amplitude and vortex phase even when they propagate to an infinite distance. This work demonstrates another state of optical vortices in free space, which will fundamentally inject new vigor into optics, and other relate scientific fields.

# Introduction

Spin and orbital angular momentum (AM) are two important characteristics of light beam. Spin AM relate to the photon spin, thereby only two states of spin AM equivalent to $\pm\hbar$ per photon are obtained by left and right circularly polarized beams, respectively. Orbital AM, however, possess unlimited states. Light beams with optical vortices $\exp(il\varphi)$ can induce orbital AM equivalent to $l\hbar$ per photon, where the topological charge $l$ is an integer number. Since the advent of orbital AM proposed in 1992 [1], light beam carrying orbital AM, namely vortex beam, has raised the upsurge of fundamental and applied researches in optics, including spin and orbital AM conversion [2, 3], rotational Doppler effect [4, 5], advanced laser [6-8], optical tweezers [9, 10], orbital AM optical communication [11-13].

Vortex beams are the basis of this new optical era. Generally, a whole optical vortices $\exp(il\varphi)$ always possesses an integer topological charge $l$. Light beams with integer optical vortices are the special solutions of wave function. That is, vortex beams, e.g. the Laguerre-Gaussian beams and Bessel-Gaussian beam, are stable states in physics that can propagate in free space [14]. For over 30 years starting from the birth of optical vortices [15], no vortex beam carrying non-integer orbital AM can stably exist in free space. Berry et. al. have already demonstrated that light beams with factional strength orbital AM cannot propagate [16-18]. Because those light beams cannot maintain the vortex phase stably like their counterparts with integer topological charge during propagating in free space. That is why, most of researches regarding orbital AM are in the framework of vortex beams with integer orbital AM [19-23].

Here, we invert this scientific consensus by proposing light beams carrying natural non-integer orbital AM in this paper. As a special solution of wave function, light beams carrying non-integer orbital AM possess a unique polarization state and an optical vortex with the topological charge $l$+0.5, where $l$ is an integer. During propagating in free space, light beams with these peculiar optical vortices can maintain their amplitudes, vortex phases with non-integer topological charge stably. This work demonstrates the propagation invariance of optical vortices with non-integer topological charge that may fundamentally facilitate extensive developments in optics, and other relate scientific fields.

# Results

In principle, light beam with optical vortices can induce a phase singularity in the center, while polarization singularity is obtained by vortex vector beam (VVB) [6]. Phase singularity represents phase uncertainty caused by a vortex phase, while polarization singularity refers to the polarization uncertainty attributed to the special polarization distribution of VVB. In the past, we tend to deem that both of them are two independent topics in free space. In 2005, Berry found a special case in a particular plane of crystal that polarization and phase singularity are interconnected, thereby giving rise to conical refraction beams with only 0.5 topological charge [24, 25]. Although it is just a special state in the crystal, this interaction inspires us that there may be optical vortices with natural non-integer topological charge in free space.

Generally, only light beams satisfied the wave function can naturally propagate in free space even to an infinite distance, such as the Laguerre-Gaussian beams and Bessel-Gaussian beam with integer orbital AM [14]. As far as we know, light beams with non-integer orbital AM have not been demonstrated to be the exact solutions of wave function. That is, one cannot assert optical vortices with non-integer topological charge are stable states in physics like those of integer one. In order to demonstrate the natural existence of light beam carrying non-integer orbital AM strictly in free space, we derive from polarization singularity. Theoretically, an $n$-order VVB with polarization directions of $\beta$ can be expressed as [6]

$$\mathbf{E}_m = [\cos(n\varphi+\beta) \quad \sin(n\varphi+\beta)]^t, \tag{1}$$

where $\varphi$ is the azimuthal angle. $t$ denotes the matrix transpose operator. Equation (1) can further be simplified into

$$\mathbf{E}_{mc} = \exp[i(n\varphi+\beta)]|\mathbf{R}\rangle + \exp[-i(n\varphi+\beta)]|\mathbf{L}\rangle, \tag{2}$$

Here, $|\mathbf{L}\rangle = [1 \quad i]^t$ and $|\mathbf{R}\rangle = [1 \quad -i]^t$ denote the left and right circularly polarized modes, respectively. From Equation (2), polarization singularity caused by $n$-order VVB can be considered as the combination of two inverse phase singularity, which are corresponding with left and right circularly polarized vortex beam with optical vortices $\exp(\pm in\varphi)$, respectively.

Supposed $n = m + \sigma$ with an integer $m$ and a fraction σ, Equation (1) represents a fractional order VVB. Vortex beams with fractional orbital AM cannot propagate in free space, thereby neither fractional order VVB. No doubt that fractional polarization singularity caused by fractional order VVB is not a stable state in physics. However, the interaction between optical vortices and VVB brings the hope of transforming fractional polarization singularity into a stable one in free space. Specifically, when modulating by an optical vortices with $\exp[i(l+\eta)\varphi]$, fractional order VVB turns into

$$\mathbf{E}_{mf} = \exp\left[i(m+\sigma+l+\eta)\varphi\right]|\mathbf{R}\rangle + \exp\left[-i(m+\sigma-l-\eta)\varphi\right]|\mathbf{L}\rangle, \quad (3)$$

where $l, m = 0, \pm 1, \pm 2...$ are integers and $0 \leq \eta \leq 1$, $0 \leq \sigma \leq 1$, respectively.

As we discussed above, only vortex beams with integer topological charge can propagate in free space. That is, the topological charge $m+\sigma+l+\eta$ and $m+\sigma-l-\eta$ must be an integer so that the overall fractional order VVB can propagate in free space. Here, we ignore $m+l$ and $m-l$ because $l, m$ are integers. In case of $0 \leq \eta \leq 1$, $0 \leq \sigma \leq 1$, one can easily obtain

$$\sigma + \eta = 1 \quad (4)$$

$$\sigma - \eta = 0 \quad (5)$$

According to Equations (4) and (5), $\sigma$=0.5 and $\eta$=0.5. Finally, the overall fractional order VVB turns into the solutions of wave function with the aid of optical vortices, which can be expressed as

$$\mathbf{E}_{mf} = \exp[i(l+0.5)\varphi]\begin{bmatrix}\cos[(m+0.5)\varphi+\beta]\\ \sin[(m+0.5)\varphi+\beta]\end{bmatrix}, \quad (6)$$

**Critical conditions for propagation invariance**

Although Equation (6) is the solution of wave function, we still cannot assert that a stable light beam is obtained by Equation (6). As we have already demonstrated in ref [26], the modulation symmetry of $n$-order VVB is broken after modulating by a vortex phase, thereby leading to two inherent circular polarization modes in Eq. (3) with different topological charge. Theoretically, once $\tau = |m+l+1| - |m-l|$ increases to a certain value, both inherent circular polarization modes in Eq. (3) are disentangled and propagate coaxially in free space. Normally, the polarization modes with higher topological charge are located at the outer ring, while the polarization modes with smaller topological charge are located at the inner ring [24]. In this case, Equation (6) cannot stand for a stable light beam.

Generally, a light beam not only is the solution of wave function, but also remains stable when propagating in free space. That is, both circular polarization modes in Eq. (3) are always intertwined together, thereby maintaining the amplitude, optical vortices $\exp[i(l+0.5)\varphi]$ in Equation (6) in free space. For this reason, $\tau = 1$, namely the smallest value of $\tau$, and the light beam in Equation (6) can propagate even to an infinite distance [see Supplementary Note 2]. It should be emphasized that both inherent circular polarization modes in Eq. (3) will be spatially separated in the focal region of a lens when $\tau > 1$, and Equation (6) cannot considered as a single light beam in the viewpoint of physics [24].

Here, one can obtain $\tau = 1$ with $l = 0$ or $m = 0$, respectively. Therefore, Equation (6) can be simplified into

$$\mathbf{E}_{mf1} = \exp(i0.5\varphi)\begin{bmatrix} \cos[(m+0.5)\varphi+\beta] \\ \sin[(m+0.5)\varphi+\beta] \end{bmatrix}, \quad (7)$$

$$\mathbf{E}_{mf2} = \exp[i(l+0.5)\varphi]\begin{bmatrix} \cos(0.5\varphi+\beta) \\ \sin(0.5\varphi+\beta) \end{bmatrix}, \quad (8)$$

Equations (7) and (8) present two different vortex beams in free space. One is a vortex beam indicated by Eq. (7) that possesses an optical vortex with the topological charge 0.5 along with a polarization state of *m*+0.5 order VVB; another is a vortex beam indicated by Eq. (8) that has a constant polarization state, namely 0.5 order VVB, but an optical vortex with arbitrary topological charge *l*+0.5. Here, the special state in the crystal are only one special case of vortex beam with *l*=0 [24, 25]. In principle, fractional order VVB cannot induce orbital AM, therefore, the whole vortex beams in Eq. (7) and (8) are naturally carrying non-integer orbital AM that roots from the optical vortices with the topological charge 0.5 and *l*+0.5, respectively.

**Vortex beam with *m*+0.5 order VVB**

Light beam in Eq. (7) possesses not only a polarization state of *m*+0.5 order VVB, but also an identical optical vortex with topological charge 0.5. Thus, to generate this vortex beam with the polarization of *m*+0.5 order VVB, one must control the polarization and phase of light beam simultaneously. For this reason, polarized spatial light modulator (SLM) developed in our previous work is utilized to generate vortex beam with the polarization of *m*+0.5 order VVB [27]. The principle of polarized-SLM are explained in Supplementary Note 1.

Figures 1 (b, g, l, q and v) present the experimental results of vortex beams with the polarization of *m*+0.5 order VVB, where (b) *m*=2; (g) *m*=3; (l) *m*=4; (q) *m*=5; (v) *m*=6, respectively. Their corresponding phases coded in the polarized-SLM are shown in Figures 1 (a, f, k, p and u), respectively. As shown in Supplementary Figure 2, these phases contain two important information of vortex beams in Figures 1(b, g, l, q and v). One is the binary phase that can be transformed into the polarization state of *m*+0.5 order VVB directly; another is a vortex phase with the topological charge of 0.5, which induces orbital AM of light beam. After passing through a polarizer (purple arrow), vortex beams with the polarization state of *m*+0.5 order VVB give rise to 2*m*+1 number of the petal-like patterns, and are rotated along with the polarizer, as shown in Figures 1 (c-e, h-j, m-o, r-t, w-y). Note that the parameters of Figures 1 (a, f, k, p and u) can be found in Supplementary Note 1.

**Vortex beam with *l*+0.5 topological charge**

For higher topological charge, one should refer to vortex beam in Equation (8). Light beams in Equation (8) possess one identical polarization, namely 0.5 order VVB, however, are carrying optical vortices with an unlimited topological charge *l*+0.5. As far as we know, vortex polarizer (VP) manufactured using the Q-plate technique can transform a linearly polarized beam into a *n* order VVB [28, 29]. Thus, these vortex beams can be easily generated using a phase-only SLM couple with a customized VP, which is indicated by the green dashed box in Supplementary Figure 3. Specifically, a collimated incident x linearly polarized beam with a wavelength of 633 nm passes through a phase-only spatial light modulator (SLM) and two lenses ($L_3$, $L_4$) and is converted into an 0.5 order VVB by a VP. Here, the VP is conjugated with the phase-only SLM by the 4f system with $L_3$ ($f_3$=150 mm) and $L_4$ ($f_4$=150 mm). That is, the fractional topological charge of vortex beam in Equation (8) can be adjusted at will by the phase-only SLM in Supplementary Figure 3.

Figures 2 (b, g, l, q and v) presents the experimental results of vortex beam with fractional topological charge *l*+0.5, which possess an identical polarization of 0.5 order VVB. Therefore, only one petal can be obtained when passing through a polarizer indicated by the purple arrow, as shown in Figures 2 (c, d, h, i, m, n, r, s, w, x). Although the light intensities in Figure 2 are similar, their vortex phases are of different topological charge. Here, we distinguish the topological charge of vortex beams using a common interferometric system. As shown in Supplementary Figure 3, the vortex beam with the polarization of 0.5 order VVB at point A transforms back into a vortex beam with x linear polarization at point B using a second VP. After passing through a BS (beam splitter), vortex beam with fractional topological charge is overlapping with a reference beam with linear polarization, thereby giving rise to a fork pattern at Point D. By observing the light intensities of fork patterns in Figures 2 (e, j, o, t and y), vortex beams in Figures 2 (b, g, l, q and v) possess optical vortices indicated by Figures 2 (a, f, k, p and u). Here, the topological charges of vortex beams in Figures 2 (b, g, l, q and v) are *l*+0.5, where (b) *l*=0; (g) *l*=1; (l) *l*=2; (q) *l*=3; (v) *l*=4, respectively. Here, their corresponding theoretical results can be found in Supplementary Figure 4.

# Discussion and conclusion

Phase and polarization singularity are two different singularities in classical optics. Phase singularity in the center of light beam are attributed to the phase uncertainty caused by the vortex phase carried by a

8vortex beam with integer orbital AM, while polarization singularity is induced by the polarization uncertainty of *n* order VVB. In theory, both singularities can remain stably individually during propagating in free space, thereby presenting two different types of light beam in optics, namely vortex beam and VVB. Unlike the above two light beams, vortex beams in Equation (7) and (8) possess a new type of singularity in the center that are the combination of a fractional phase singularity and a fractional polarization singularity. Fractional phase and polarization singularities are related to a fraction vortex phase and a fraction order VVB, respectively. Owing to the interaction of optical vortices and polarization, these vortex beams with fractional topological charge can maintain their amplitude and vortex phase even when they propagate to an infinite distance [see Supplementary Note 2]. This propagation invariance not only manifests that faction strength orbital AM is a stable state in physics that can propagate in free space like that of strength orbital AM, but also verifies two different light beams in optics, namely *m*+0.5 order VVB and vortex beam with *l*+0.5 topological charge.

In conclusion, we have demonstrated light beams carrying natural non-integer orbital AM in this paper. Their amplitude and vortex phases with fractional topological charge maintain stably when propagating in free space. The advent of these peculiar vortex beams will not only deepen our understanding about optical vortices, but also fundamentally inject new vigor into optics, and other relate scientific fields by providing an additional choice for researchers in the field of optical vortices.

## Data Availability

All data supporting the findings of this study are available from the corresponding author on request.

**Acknowledgments**

Parts of this work were supported by the National Natural Science Foundation of China (62022059/11804232) and the National Key Research and Development Program of China (2018YFC1313803).

**Author contributions**

X. Weng conceived of the research and designed the experiments. Y. Miao, G. Wang and Q. Zhan performed the experiments. X. Dong and X. Weng analyzed all of the data and carried out the numerical simulations. X. Gao supervised the experiments. X. Weng and X. Gao co-wrote the paper with support from the other authors. J. Qu, X. Gao and S. Zhuang offered advice regarding its development. S. Zhuang directed the entire project. All of the authors participated in the analysis and discussion of the results.

**Competing interests statement**

The authors declare that they have no competing financial or nonfinancial interests to disclose.

# Figures

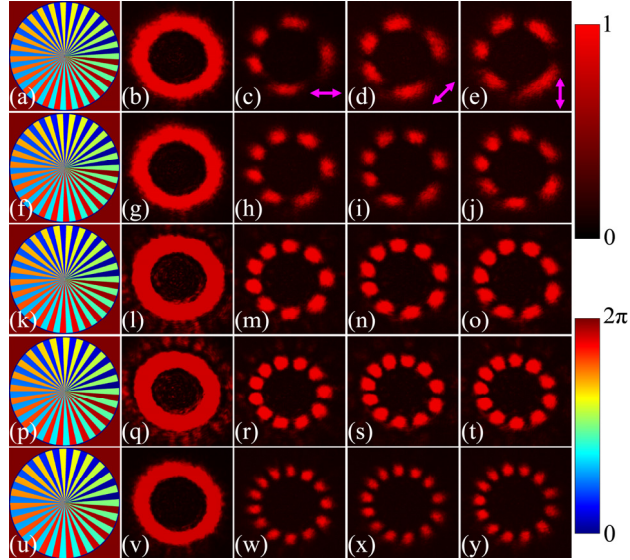

**Figure 1.** Experimental results of vortex beam with the polarization state of *m*+0.5 order VVB. Vortex beam with (b) *m*=2, (g) *m*=3, (l) *m*=4, (q) *m*=5, (v) *m*=6 are created by the phases (a, f, k, p, u), respectively. Subfigures (c-e, h-j, m-o, r-t, w-y) show the light intensities of vortex beams passing through the polarizer indicated by the purple arrow.

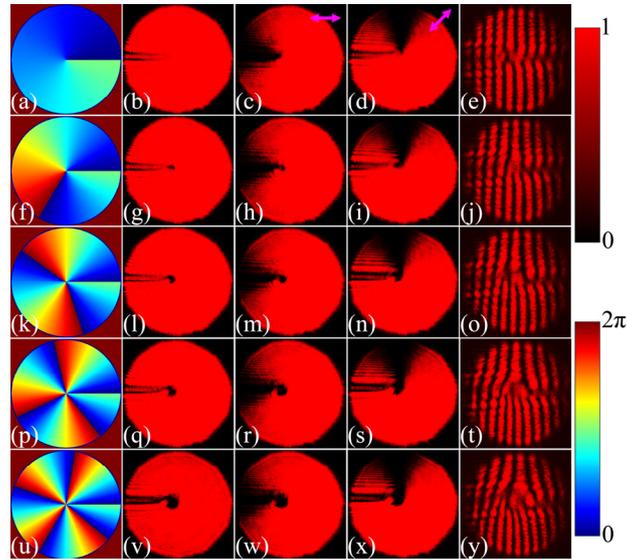

**Figure 2.** Experimental results of vortex beam with fraction topological charge *l*+0.5. Vortex beam with the polarization state of 0.5 order VVB (b, g, l, q, v) are created using a designed VP. Their corresponding vortex phases are shown in (a, f, k, p, u), where the topological charges are *l*+0.5 (a) *l*=0, (f) *l*=1, (k) *l*=2, (p) *l*=3, (u) *l*=4, respectively. Subfigures (c, d, h, i, m, n, r, s, w, x) show the light intensities of vortex beams passing through the polarizer indicated by the purple arrow, and (e, j, o, t, y) are the fork patterns formed by overlapping with a reference beam in Supplementary Figure 3.

## Supplementary Materials for

## Light beam carrying natural non-integer orbital angular momentum in free space

Xiaoyu Weng, Yu Miao, Guanxue Wang, Qiufang Zhan, Xiangmei Dong, Junle Qu, Xiumin Gao and Songlin Zhuang

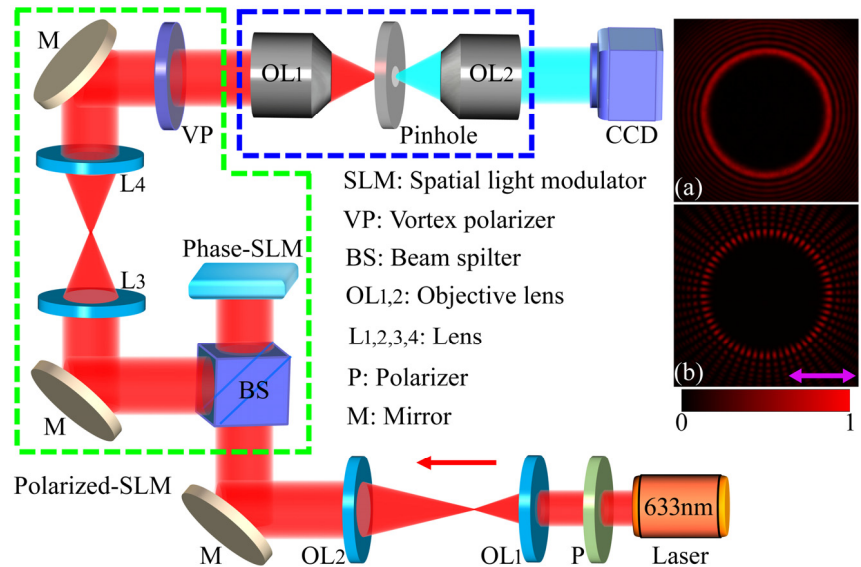

**Supplementary Figure 1** Polarized-SLM based on phase vectorization. The phase control system (green dot box) and filter system (blue dot box) compose an entire polarized-SLM. A collimated incident *x* linearly polarized beam with wavelength 633nm propagating along the optical axis passes through a phase-only SLM and two lenses ($L_3$, $L_4$) before it is converted into a $\delta=30$ order VVB by vortex polarizer (VP). VP can be manufactured using the technique of Q-plate. $L_3$ ($f_3$=150 mm) and $L_4$ ($f_4$=150 mm) compose a 4f-system that makes the phase coded in the phase-only SLM and VP conjugate. (a, b) are the light intensities of $\delta=30$ order VVB without and with a polarizer (purple arrow), respectively. The modulated VVB output from the first system is filtered by the filter system so that the fundamental phase-to-polarization link is established. By adjusting the phase coded in phase-only SLM, one can achieve pixelate polarization modulation of light beam dynamically, which is recorded using a CCD. Here, both numerical apertures (NA) of OL1, OL2 are 0.01, and the red arrow represents the propagation direction of light beam.

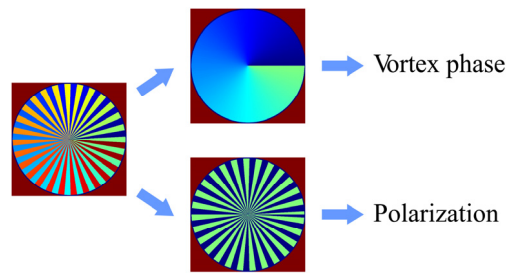

**Supplementary Figure 2** Schematic of the overall phase for creating vortex beam in Equation (7). The phases coded in polarized-SLM shown in Supplementary Figure 1 can be divided into two parts: one is vortex phase that indicates the optical vortex carried by the light beam; another is the binary phase that is responsible for generating the polarization of *m*+0.5 order VVB.

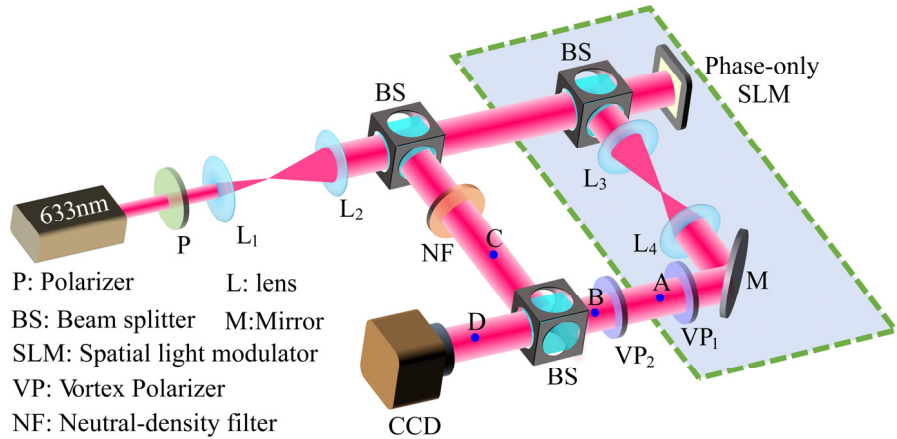

**Supplementary Figure 3** Schematic of an interferometric system. An incident linearly polarized beam divides into a reference beam and a measurement beam using a BS (beam splitter). After passing through the vortex beam creation system formed by a phase-only SLM couple with a VP (green dot box), the measurement beam turns into a vortex beam in Equation (8) at point A. The vortex beam at point A eventually transforms back into a vortex beam with linear polarization by the second VP at point B. When overlapping with the reference beam using the BS, fork pattern can be obtained at point D and recorded by the CCD. Note that the light intensity of reference beam at point C can be adjusted to be equal with that of light beam at point B using a NF.

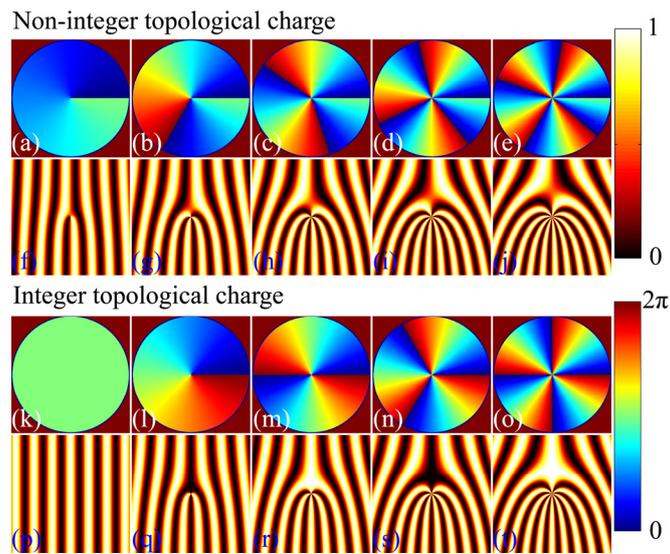

**Supplementary Figure 4** Theoretical result of fork pattern. By overlapping with the reference beam, different topological charge of vortex beam gives rise to different fork pattern. Here, the topological charges of vortex phase in (a-e) are (a) 0.5; (b) 1.5; (c) 2.5; (d) 3.5; (e) 4.5, while those of vortex phase in (k-o) are (k) 0; (l) 1; (m) 2; (n) 3; (o) 4. Their corresponding fork patterns are shown in (f-j) and (p-t), respectively.

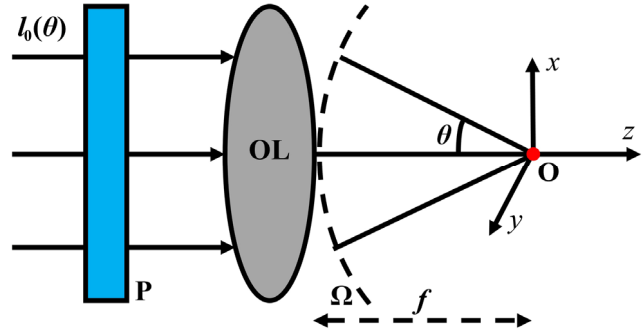

**Supplementary Figure 5** Schematic of the focusing system. $\Omega$ is the focal sphere, with its center at O and radius $f$, namely, the focal length of the objective lens (OL). The transmittance of the pupil filter P implies the phase of vortex beam carrying non-integer orbital AM in Equations (7) and (8). $\theta$ is the convergent angle. $l_0(\theta)$ denotes the electric field amplitude of the incident vortex beam.

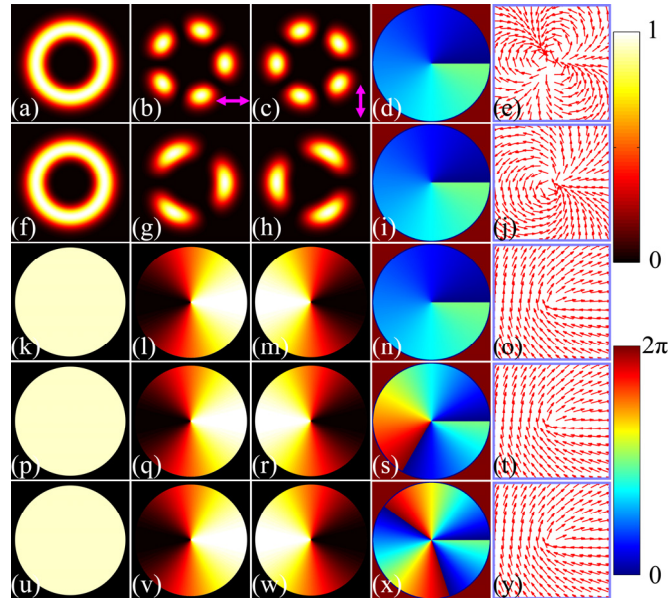

**Supplementary Figure 6.** Vortex beams at the wavefront of OL. The order and topological charge of vortex beam with $\beta = 0$ in Equations (7) and (8) are $m+0.5$ and $l+0.5$, respectively. Here, (a) $m=2$, $l=0$; (f) $m=1$, $l=0$; (k) $m=0$, $l=0$; (p) $m=0$, $l=1$; (u) $m=0$, $l=2$. Their polarization states and vortex phases are shown in (d, i, n, s, x) and (e, j, o, t, y), respectively. When passing through the polarizer indicated by the purple arrow, petals with $2m+1$ number are obtained and rotated along with the polarizer, see (b, c, g, h, l, m, q, r, v, and w).

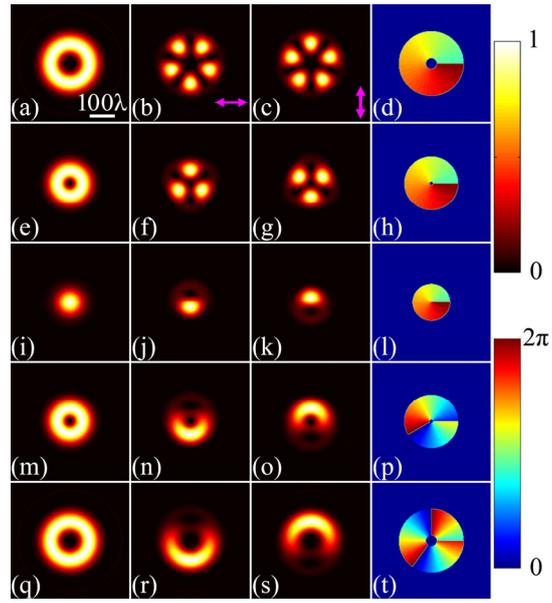

**Supplementary Figure 7.** Vortex beams in the focal region of OL. Subfigures (a, e, i, m, q) are the corresponding vortex beams of Supplementary Figure 6 in the focal region of OL. Their vortex phases are shown in (d, h, l, p, t), respectively. When passing through the polarizer indicated by the purple arrow, petals with $2m+1$ number are obtained and rotated along with the polarizer, see (b, c, f, g, j, k, n, o, r, s).

# Supplementary Note 1: Generation of vortex beam with non-integer topological charge

## Part 1: Polarized-SLM based on phase vectorization

Generation of vortex beam with the polarization of $m+0.5$ order VVB requires the polarization and phase modulation of incident light beam. Therefore, we utilize the polarized-SLM (spatial light modulator) developed in our previous work [1], which can control both properties of light beam simultaneously and dynamically. Supplementary Figure 1 presents the schematic of polarized-SLM based on phase vectorization, which is compose with a phase control system (green dot box) and a filter system (blue dot box). A collimated incident x linearly polarized beam with wavelength 633nm propagating along the optical axis is converted into a $\delta=30$ order VVB by VP. VP is a vortex polarizer realized by the Q-plate technique [2, 3]. The phase of $\delta=30$ order VVB can be adjusted at will by the phase-only SLM, which are conjugated with VP by the 4f system formed by $L_3$ ($f_3$=150 mm) and $L_4$ ($f_4$=150 mm). The modulated $\delta=30$ order VVB output from the first system are further filtered by the filter system using a pinhole with radius 400μm. Finally, the phase of $\delta=30$ order VVB in Supplementary Equation (1) can link with its polarization in Supplementary Equation (2) directly, and the entirely polarized-SLM is built.

$$\phi=\text{Phase}[\cos\varphi_0 \exp[i(\delta\varphi-\eta)]+\sin\varphi_0 \exp[-i(\delta\varphi-\eta)]]. \tag{1}$$

$$\mathbf{E}=\cos\varphi_0 \exp(-i\eta)|\mathbf{L}\rangle+\sin\varphi_0 \exp(i\eta)|\mathbf{R}\rangle. \tag{2}$$

Here, $|\mathbf{L}\rangle=[1 \quad i]^t$ and $|\mathbf{R}\rangle=[1 \quad -i]^t$ denote the left and right circular polarization modes, respectively. $t$ denotes the matrix transpose operator. Both numerical apertures (NA) of $OL_1$, $OL_2$ are 0.01, and the red arrow represents the propagation direction of light beam.

Suppose an additional phase $\psi$ is superposed with the phase $\phi$ in Supplementary Equation (1). Then, the entire VVB phase can be expressed as $\Omega=\phi+\psi$. The phase $\phi$ and $\psi$ represent two functions of polarized-SLM. For the first term, $\phi$ represents three particular phases: binary phase, vortex phases, and combinations of both, which are corresponding to linear, circular, and elliptical polarization, respectively. The second term $\psi$ represents the pure phase modulation of the light beam. Accordingly, the overall phase of creating vortex beam with fractional topological charge can be divided into two parts, as shown in Supplementary Figure 2. One is the binary phase related to the polarization of $m+0.5$ order VVB in Equation (7), another is the vortex phase with fractional topological charge, namely $\psi = 0.5\varphi$.

**Parameter of vortex beams in Figure 1**

According to Supplementary Equation (1), the phase for linear polarization can be simplified to

$$\phi=\text{Phase}\{\cos[(\delta\varphi-\eta)]\}, \tag{3}$$

where $\delta = 30$ is the order of the VVB; $\eta$ denotes the angle between the polarization direction and the x axis. Therefore, the linear polarization output from polarized-SLM can be expressed as

$$\mathbf{E} = \cos\eta|\mathbf{x}\rangle + \sin\eta|\mathbf{y}\rangle, \tag{4}$$

where $|\mathbf{x}\rangle=[1\ 0]^t$ and $|\mathbf{y}\rangle=[0\ 1]^t$ denote the x and y linearly polarized modes, respectively. That is, one can simply obtain the polarization of fraction order VVB by the parameter $\eta=(m+0.5)\varphi$. By overlapping a vortex phase with 0.5 topological charge, namely $\psi=0.5\varphi$, the overall phases for the creation of vortex beams in Figures 1(a, f, k, p, u) can be simplified into $\Omega=\phi+\psi$, where (a) $m=2$; (f) $m=3$; (k) $m=4$; (p) $m=5$; (u) $m=6$, respectively.

**Part 2: Vortex beam with $l+0.5$ topological charge**

Supplementary Figure 3 presents an interferometric system, which is commonly utilized to demonstrate the phase of vortex beam with $l+0.5$ topological charge in Equation (8). An incident linearly polarized beam divides into a reference beam and a measurement beam using a BS (beam splitter). After passing through the vortex beam creation system formed by a phase-only SLM couple with a VP, the light beam at point A can be expressed as

$$\mathbf{E}_v = \exp[i\psi]M_{vp}\mathbf{E}_i, \tag{5}$$

Here, the VP is conjugated with the phase-only SLM by the 4f system with $L_3$ ($f_3=150$ mm) and $L_4$ ($f_4=150$ mm). $\psi = (l+0.5)\varphi$ is the phase coded in the phase-only SLM and $\mathbf{E}_i = [1\ 0]^t$ denotes the polarization of measurement beam. $M_{vp}$ indicates the Jones matrix of VP, which can be written as [4]

$$M_{vp} = \begin{bmatrix} \cos(m\varphi+\varphi_0) & \sin(m\varphi+\varphi_0) \\ \sin(m\varphi+\varphi_0) & -\cos(m\varphi+\varphi_0) \end{bmatrix}, \tag{6}$$

Therefore, the vortex beams in Equation (8) are obtained at point A with $m=0.5$ and $\varphi_0=0$. Their corresponding light intensities are shown in Figure 2.

The vortex beam at point A further propagates along the optical axis and transforms back into a

vortex beam with x linear polarization by a second VP. That is, the light beam at point B can be expressed as

$$\mathbf{E}_B = M_{vp}\mathbf{E}_v, \qquad (7)$$

which can further be simplified into $\mathbf{E}_B = \exp[i(l+0.5)\varphi][1\ \ 0]^t$.

When overlapping with the reference beam at point C, different topological charge of vortex beam at point B induces different light intensity of fork pattern at point D, thereby verifying the vortex beam carrying non-integer orbital AM in optics. Supplementary Figure 4 presents the theoretical results of fork pattern created by the superposition of reference beam and vortex beam with integer and non-integer topological charge, where the topological charge of vortex phases in Supplementary Figure 4 are (a-e) 0.5-4.5; (k-o) 0-4, respectively. Theirs corresponding fork patterns are shown in Supplementary Figure 4 (f-j) and (p-t), respectively. According to these fork pattern, one can simply distinguish vortex phase with different topological charge. Note that the light intensity of reference beam can be adjusted to be equal with that of light beam at point B using a NF (Neutral-density filter).

# Supplementary Note 2: Propagation invariance of vortex beam carrying non-integer orbital AM

In this note, we demonstrate the propagation invariance of vortex beam carrying non-integer orbital AM. Similar to the vortex beam carrying integer orbital AM, light beams in Equations (7) and (8) should maintain their optical vortices invariance during propagating in free space, which are collectively called propagation invariance of light beam.

To verify the propagation invariance of vortex beam carrying non-integer orbital AM, we compare the polarization, vortex phase and amplitude of light beam at the wavefront with their counterparts after propagating to an infinite distance. According to information optics, one can simply obtain the image of light beam at an infinite distance in the focal region of an objective lens (OL). As shown in Supplementary Figure 5, the electric fields of vortex beam in the focal region of the OL can be expressed as [5]

$$\mathbf{E} = -\frac{A}{\pi} \int_0^{2\pi} \int_0^{\alpha} \sin\theta \cos^{1/2}\theta T l_0(\theta) \mathbf{V} \exp(-i k \mathbf{s} \cdot \boldsymbol{\rho}) d\theta d\varphi, \quad (8)$$

Here, $A$ is a normalized constant. $\theta$ and $\varphi$ are the convergent and azimuthal angle, respectively. $\alpha = \arcsin(\mathrm{NA}/\upsilon)$, where NA is the numerical aperture of the OL and $\upsilon$ is the refractive index in the focusing space. $\boldsymbol{\rho} = (r\cos\phi, r\sin\phi, z)$ denotes the position vector of an arbitrary field point. The unit vector along a ray is expressed as $\mathbf{s} = (-\sin\theta\cos\varphi, -\sin\theta\sin\varphi, \cos\theta)$. The wavenumber is $k = 2\upsilon\pi/\lambda$ with the wavelength of the incident beam $\lambda$. $T = \exp(i\psi)$ represents the transmittance of the pupil filter P, where $\psi = (l+0.5)\varphi$ denotes the vortex phase of the vortex beams in Equations (7) and (8). $l_0(\theta)$ is the electric field amplitude of the incident beam, which can be expressed as [6]

$$l_0(\theta) = J_1(2\beta_0 \frac{\sin\theta}{\sin\alpha}) \exp[-(\beta_0 \frac{\sin\theta}{\sin\alpha})^2], \quad (9)$$

where $\beta_0$ is the ratio of the pupil radius to the incident beam waist. $J_1(\bullet)$ is the Bessel function of the first kind with order 1.

According to Equation (6), incident vortex beam possesses the polarization state of ($m$+0.5) order VVB, which can be considered as the combination of x and y linear polarization modes. Therefore, in Supplementary Equation (8), the propagation unit vector of the incident beam immediately after having passed through the lens is $\mathbf{V} = \cos[(m+0.5)\varphi]\mathbf{V}_x + \sin[(m+0.5)\varphi]\mathbf{V}_y$, where $\mathbf{V}_x$ and $\mathbf{V}_y$ are the electric vectors of $|\mathbf{x}\rangle$ and $|\mathbf{y}\rangle$, respectively, and can be written as [1]

$$\mathbf{V}_x = \begin{bmatrix} \cos\theta + (1-\cos\theta)\sin^2\varphi \\ -(1-\cos\theta)\sin\varphi\cos\varphi \\ \sin\theta\cos\varphi \end{bmatrix}; \mathbf{V}_y = \begin{bmatrix} -(1-\cos\theta)\sin\varphi\cos\varphi \\ 1-(1-\cos\theta)\sin^2\varphi \\ \sin\theta\sin\varphi \end{bmatrix}. \qquad (10)$$

Eventually, the focal light intensity of the vortex beam in Equations (7) and (8) can be obtained using $I = |\mathbf{E}|^2$.

**Comparison of vortex beam at the wavefront and infinite distance**

In the following simulations, NA=0.01, $\upsilon = 1$, and $\beta_0 = 1$. The unit of length in all figures is the wavelength $\lambda$, and the light intensity is normalized to the unit value.

Supplementary Figure 6 presents vortex beams in Equations (7) and (8) at the wavefront of OL. Here, the order and topological charge of vortex beams with $\beta = 0$ in Supplementary Figure 6 are $m+0.5$ and $l+0.5$, respectively, where (a) $m$=2, $l$=0; (f) $m$=1, $l$=0; (k) $m$=0, $l$=0; (p) $m$=0, $l$=1; (u) $m$=0, $l$=2, respectively. Their polarization states and vortex phases are shown in Supplementary Figure 6 (d, i, n, s, x) and Supplementary Figure 6 (e, j, o, t, y), respectively. When passing through the polarizer indicated by the purple arrow, petals with $2m+1$ number are obtained and rotated along with the polarizer, see Supplementary Figure 6 (b, c, g, h, l, m, q, r, v, and w). Because of the natural optical vortices with topological charge $l$+0.5, the energy fluxes of vortex beams are rotating continuously during propagating in free space. Therefore, when propagating into an infinite distance in free space, the overall polarization state of light beam are transformed into a $m$+0.5 order VVB with $\beta = 0.25\pi$, as shown in Supplementary Figure 7 (a-c, e-g, i-k, m-o, q-s). Despite of this polarization rotation, the amplitude and vortex phase of light beam maintain [see Supplementary Figure 7 (a, e, i, m, q) and (d, h, l, q, t)], thereby verifying that optical vortices with non-integer topological charge naturally exist in physics.

## Supplementary References